\documentclass[aps,prc,twocolumn,groupedaddress,showpacs]{revtex4}
\usepackage[dvips]{graphicx}
\usepackage{amsmath}

\begin{document}
\title{Ordered bicontinuous double-diamond morphology in subsaturation nuclear matter}

\author{Masayuki Matsuzaki}
\email[]{matsuza@fukuoka-edu.ac.jp}
\affiliation{Department of Physics, Graduate School of Sciences, Kyushu University,
             Fukuoka 812-8581, Japan}

\affiliation{Department of Physics, Fukuoka University of Education, 
             Munakata, Fukuoka 811-4192, Japan}\thanks{permanent address}

%\date{\today}

\begin{abstract}
 We propose to identify the new ``intermediate" morphology in subsaturation nuclear 
matter observed in a recent quantum molecular dynamics simulation with the ordered 
bicontinuous double-diamond structure known in block copolymers. We estimate its 
energy density by incorporating the normalized area-volume relation given in a 
literature into the nuclear liquid drop model. The resulting energy density is 
higher than the other five known morphologies. 
\end{abstract}

\maketitle

 Understanding the form of existence of nuclear matter in extreme environments is 
of importance both from the point of view of nuclear many body problem and from the 
context of nuclear astrophysics. Nuclear matter with subsaturation densities 
(0.1$n_\mathrm{s}$ -- $n_\mathrm{s}$; $n_\mathrm{s}$ being the saturation density) 
is believed to exist in the inner crust of neutron stars and to appear at stellar 
collapses. This density range corresponds to the transitional region between a 
Coulomb lattice of spherical nuclei and a uniform matter; Ravenhall et al.~\cite{rav} 
and Hashimoto et al.~\cite{hash} showed that nuclear matter experiences various 
phases in the course of density change. After that, this was elucidated by various 
model calculations~\cite{oya1,wil,las,lor,oya2,wat1,iid}. With the help of the 
recent progress of computer power, not only the ground state at each density but also 
the dynamical phase transitions between them and the excited states were studied by 
means of first-principle numerical simulations with the quantum molecular dynamics 
(QMD) method~\cite{mar1,wat2,wat3,hor}. Among them, the most basic 
is to understand the form of existence of nuclear matter at each density. 
Watanabe et al.~\cite{wat2} found, in addition to the five known morphologies 
--- sphere, cylinder, slab, cylindrical hole (tube), and spherical hole (bubble) 
---, a new ``intermediate" morphology that is characterized by negative Euler 
characteristic, $\chi < 0$. They described this as a highly connected spongelike 
shape, and conjectured its relevance to astrophysics. 

 However, these five morphologies were known~\cite{ino} in block copolymers 
of, for example, styrene and isoprene. In the nuclear case, the two domains are 
composed of matter and void (or very dilute neutron vapor), whereas they are 
composed of two kinds of polymers such as polystyrene (PS) and polyisoprene (PI) 
in the macromolecule case. 
Their morphologies are determined by a balance between the surface energy and the 
Coulomb energy in the former, whereas by that between the (inter-)surface energy 
and the stretching free energy, like that causes rubber elasticity, in the latter. 
A new morphology was found experimentally in a star block copolymer by Thomas et 
al.~\cite{tho} and in a diblock copolymer by Hasegawa et al.~\cite{hase} between 
the PS cylinder and the lamella (slab) and between the lamella and the PI cylinder. 
These experiment determined essentially the correct morphology 
(Fig.~4 in Ref.~\cite{tho} and Fig.~3 in Ref.~\cite{hase}). 
Soon thereafter the shape of the interface between the two microphases was 
mathematically recognized as the $H$ surface --- a family of surfaces with constant 
mean curvature $H$~\cite{and}. The $H$ surfaces are known to minimize the area under 
the symmetry and volume conservation condition. The observed morphology is called 
the ordered bicontinuous double-diamond (OBDD) structure according to its symmetry. 
The OBDD structure consists of two interwoven networks of tetrahedral units 
(four fold junctions) filled by one material and the remaining matrix filled by 
the other.
Calculations of its free energy were done by several groups~\cite{and,lik,olm} and 
they concluded that the OBDD structure is not the ground state at any PS/PI 
composition. 

 Both from the location --- adjacent to the slab --- of the ``intermediate" phase 
of Watanabe et al. and their observation that it is highly connected spongelike, 
it looks quite natural to interpret this phase as the OBDD structure observed in 
block copolymers. A direct calculation of the energy density of this structure based 
on some microscopic nuclear Hamiltonian is desirable, but it is too much complicated 
unfortunately. Alternatively, here we estimate its energy density by combining the 
familiar liquid drop model relation and the normalized area-volume relation of the 
OBDD morphology given in the literature~\cite{and}. 

\begin{figure}[htbp]
  \includegraphics[width=7cm]{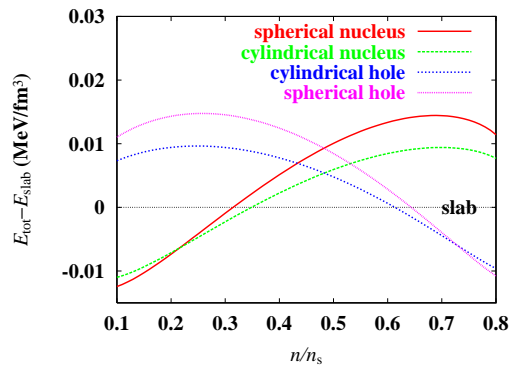}
 \caption{(Color online) Total energy density of each morphology relative to that of the 
          slab, as a function of the average density. \label{fig1}}
\end{figure}

\begin{figure}[htbp]
  \includegraphics[width=7cm]{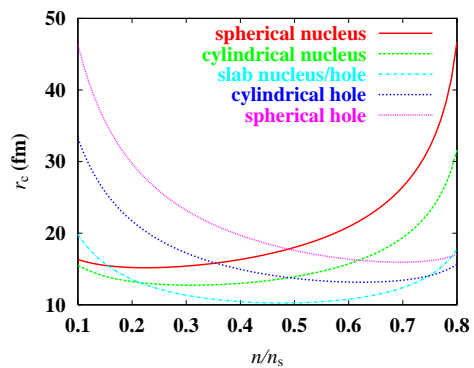}
 \caption{(Color online) Cell size of each morphology  as a function of the average density. 
          \label{fig2}}
\end{figure}

 The liquid drop model relations are taken from Ravenhall et al.~\cite{rav} who 
first predicted the non-spherical morphologies. Under the Wigner-Seitz cell 
approximation, the total energy density is given by 
$E_\mathrm{S}+E_\mathrm{C}+E_\mathrm{B}+E_e$, a sum of the surface energy, 
the Coulomb energy, the bulk energy, and the kinetic energy of electron gas. 
Here the Coulomb energy consists of the nuclear electrostatic energy and the lattice 
energy of electron gas and a spatially spread nucleus embedded in it. Zero 
temperature is assumed. The model is formulated by extending that for the spherical 
case given in Ref.~\cite{bay}. We consider spherical ($d = 3$), cylindrical 
($d = 2$), and slab ($d = 1$) cases as in Ref.~\cite{rav}. The unit cell for each 
case is a sphere with the radius $r_\mathrm{c}$, a cylinder with the radius 
$r_\mathrm{c}$ and the length $l$, and a rectangular parallelepiped with the sides 
$2 r_\mathrm{c}$, $a$, $b$, respectively. The average density over the cell is 
$n$. In each cell, nuclear matter is put in the form of a sphere with the radius 
$r$, a cylinder with the radius $r$ and the length $l$, and a rectangular 
parallelepiped with the sides $2 r$, $a$, $b$, respectively. The density of nuclear 
matter is $n'$. The other part of the cell is occupied by a very dilute neutron vapor. 
The proton fraction is $x$. The volume fraction is $u=n/n'=(r/r_\mathrm{c})^d$. 
The surface tension is $\sigma$. Among the total energy density, $E_\mathrm{S}$ 
and $E_\mathrm{C}$ depend on the shape and size of the cell and given by 
\begin{gather*}
E_\mathrm{S}=\frac{u\sigma d}{r} , \\
E_\mathrm{C}=2\pi n'^2x^2e^2r^2uf_d(u) , \\
f_d(u)=\frac{1}{d+2}\Big[\frac{2}{d-2}\Big(1-\frac{1}{2}du^{1-\frac{2}{d}}\Big)+u\Big] , \\
\lim_{d\to2}f_d(u)=\frac{1}{4}\Big(-1-\ln u+u\Big) .
\end{gather*}
The variation of $E_\mathrm{S}+E_\mathrm{C}$ with respect to $r$ gives the familiar 
relation
\begin{equation}
E_\mathrm{S}=2E_\mathrm{C} .
\label{eq1}
\end{equation}
The bulk energy is given by
\begin{equation*}
E_\mathrm{B}=un'\Big[E_0+\frac{K}{18}\Big(1-\frac{n'}{n_\mathrm{s}}\Big)^2\Big] ,
\end{equation*}
with $E_0$ and $K$ being the binding energy per baryon and the incompressibility. 
The variation of $E_\mathrm{S}+E_\mathrm{C}+E_\mathrm{B}$ with respect to $n'$ 
gives an equation that determines $u$~\cite{rav}. That for the cylindrical and 
spherical hole morphologies is similar. Thus, $n'$, $r$, and $r_\mathrm{c}$ for 
each $n$ and morphology are determined. The electron energy that is common to all 
morphologies are irrelevant to energy comparison but can be given by 
\begin{gather*}
E_e=\frac{3}{4}\hbar c\Big(3\pi^2n_e\Big)^\frac{1}{3}n_e , \\
n_e=xn ,
\end{gather*}
as an ultrarelativistic gas~\cite{bay}. 
Adopting the parameter set $x = 0.3$, $E_0 = -11.4$ MeV, $K = 291$ MeV, 
$n_\mathrm{s} =0.147$ fm$^{-3}$, and $\sigma = 0.74$ MeV/fm$^2$ given in 
Ref.~\cite{rav} and relevant to stellar collapses, the relative energy density and 
the cell size are obtained as in Figs.~\ref{fig1} and \ref{fig2}. Figure~\ref{fig1} 
indicates the sequential shape change as the density change. The cell sizes of the 
spherical nucleus and hole in Fig.~\ref{fig2} are used later. 

\begin{figure}[htbp]
  \includegraphics[width=7cm]{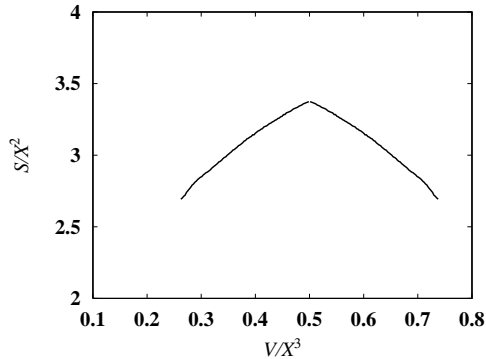}
 \caption{Normalized area-volume relation for the OBDD 
          structure. This is adapted from that for the single-diamond structure in 
          Ref.~\cite{and}. The derivative is proportional to the mean curvature. 
          \label{fig3}}
\end{figure}

 In the above model for the five known morphologies, $E_\mathrm{S}\propto r^{-1}$ 
and $E_\mathrm{C}\propto r^2$ are given independently and accordingly the variation 
of their sum with respect to $r$ leads to Eq.~(\ref{eq1}). In the case of the OBDD 
structure, however, $E_\mathrm{S}$ and $E_\mathrm{C}$ can not be represented 
simply (at least to the author's knowledge). Alternatively, we can utilize the 
non-dimensionalized area-volume relation given for a cubic cell in Ref.~\cite{and}. 
With a lattice parameter $X$, this gives $S/X^2$ as a function of $V/X^3$. 
Consequently, the surface energy density is given by 
\begin{equation*}
E_\mathrm{S}=\sigma\frac{S}{X^2}\frac{1}{X} ,
\end{equation*}
as a function of $u\equiv V/X^3$. The relation between $u$ and $n$ is taken from the 
$d = 3$ (nucleus or hole) case. Assuming that Eq.~(\ref{eq1}) holds also for this 
morphology, the Coulomb energy is automatically determined. This means that the total 
energy can be obtained since $E_\mathrm{B}$ and $E_e$ are independent of morphology. 
The area-volume relation in Fig.~\ref{fig3} for the OBDD morphology was adapted from 
Fig.~1(b) in Ref.~\cite{and} for the single-diamond structure. Since the relation 
\begin{equation*}
\frac{dS}{dV}=2H
\end{equation*}
can be derived form the first variational formula of area, the nuclear OBDD 
($V/X^3 < 0.5$) is a family of surfaces with $H > 0$, while the hole OBDD 
($V/X^3 > 0.5$) is that with $H < 0$ (see Fig.~4 in Ref.~\cite{wat2}). 
Although the $H > 0$ and $H < 0$ parts are connected smoothly in the case of 
the single-diamond structure, their curvatures are discontinuous in the case of 
the double-diamond structure. This indicates that the lamella structure exists 
between them.

\begin{figure}[tbp]
  \includegraphics[width=7cm]{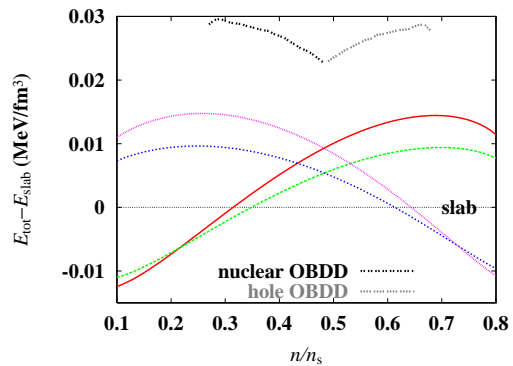}
 \caption{(Color online) Total energy density of nuclear and hole OBDD structures relative to 
           that of the slab, as functions of the average density. Those for the known 
           morphologies are the same as in Fig.~\ref{fig1}. \label{fig4}}
\end{figure}

 Assuming $X=\Big(\frac{4\pi}{3}\Big)^{1/3}r_\mathrm{c}(d=3)$ (see Fig.~\ref{fig2}), 
we estimate the energy density of the OBDD phase. The result is shown in 
Fig.~\ref{fig4}. This figure indicates that this simple estimate gives 
25 -- 30 keV/fm$^3$ higher energy for the OBDD structure than the slab. Qualitatively, 
this result is consistent with that the OBDD phase is not the ground state at any 
composition in block copolymers. In the previous works, the ``cross" phase in 
Ref.~\cite{wil} and the ``mixed" phase in Ref.~\cite{las} might correspond to the 
OBDD phase although our estimate gives higher energy. 
We did not try to change the parameters given 
in Ref.~\cite{rav} because all parameters correlate and they require Skyrme model 
calculation that is beyond the scope of the present simple estimate. 

 To summarize, we proposed to identify the new ``intermediate" morphology in subsaturation 
nuclear matter observed in a recent QMD simulation with the ordered bicontinuous 
double-diamond structure known in block copolymers. We estimated its energy density in a 
hybrid manner --- incorporating the normalized area-volume relation given mathematically 
in a literature into the nuclear liquid drop model. The resulting energy density is higher 
than the other five known morphologies; this is qualitatively consistent with the results 
for block copolymers. 

 {\it Note added in proof}

\noindent
 After submission of the manuscript, the author found that another 
constant mean curvature surface, the bicontinuous double-gyroid 
structure similar to the OBDD but consisting of three-fold junctions~\cite{haj}, 
is favored in block copolymers. In the nuclear case, mathematical 
surfaces are meaningful as an idealization and therefore it would be 
difficult to distinguish them. 

 The author thanks S. Ei for directing his interest to block copolymers.

\end{document}